# Charge Transfer and Functionalization of Monolayer InSe by Physisorption of Small Molecules for Gas Sensing

Yongqing Cai, Gang Zhang* and Yong-Wei Zhang*

*Institute of High Performance Computing, A*STAR, Singapore 138632*

## ABSTRACT

First-principles calculations are performed to investigate the effects of the adsorption of gas molecules ($CO$, $NO$, $NO_2$, $H_2S$, $N_2$, $H_2O$, $O_2$, $NH_3$ and $H_2$) on the electronic properties of atomically thin indium selenium (InSe). Our study shows that the lone-pair states of Se are located at the top of the valence band of InSe and close to the Fermi energy level, implying its high sensitivity to external adsorbates. Among these gas molecules, $H_2$ and $H_2S$ are strong donors, $NO$, $NO_2$, $H_2O$ and $NH_3$ are effective acceptors, while $CO$ and $N_2$ exhibit negligible charge transfer. The $O_2$ molecule has very limited oxidizing ability and a relatively weak interaction with InSe which is comparable to the $N_2$ adsorption. A clear band gap narrowing is found for the $H_2S$, $NO_2$, and $NH_3$ adsorbed systems whereas a Fermi level shifting to the conduction band is observed upon a moderate uptake of $H_2$ molecules. Our analysis suggests several interesting applications of InSe: 1). Due to the different interaction behaviors with these external molecules, InSe can be used for gas sensing applications; 2). by monitoring the adsorption/desorption behavior of these gas molecules, the population of hole states in InSe due to photon stimulation or defect production can be quantitatively estimated; and 3). it is promising for novel electronic and optoelectronic applications since the adsorption-induced in-gap states and strong charge transfer are able to change the content and polarity of charged carriers and lead to different optical properties.

I. INTRODUCTION

Two-dimensional (2D) materials have received a great deal of attention in recent years due to their intriguing electronic properties originated from quantum confinement and high surface-volume ratio[1-6]. In particular, layered 2D materials with van der Waals (vdW) interaction between atomically thin layers allow for high-quality exfoliation/transfer and ultrahigh flexibility, and thus are highly promising for flexible nanoelectronics applications [7-10]. For 2D semiconducting materials with a finite band gap like $MoS_2$ and phosphorene, their electronic band gaps and work functions are generally layer number-dependent [11], which allows for more efficient solar utilization and carrier injection at the 2D materials-electrodes interfaces. In addition, their electronic properties and structural stability are sensitive to environmental molecules due to the high surface-volume ratio and weak electronic screening [12-16]. As a result, 2D materials are highly suitable for gas sensing via modulating the carrier density and shifting the Fermi level [12, 13]. Last but not least, 2D materials can be easily functionalized by molecular doping, which can be used to modify their electronic, optical, and thermoelectric properties [17-23], in addition to strain engineering and heterogeneous construction [24, 25].

More recently, indium selenides (InSe), a layered material with each InSe layer being composed of a Se–In–In–Se structure, gains increasing attention [26-51]. Many studies have been performed to understand its charge carrier dynamics [26,27,38], structure stability [28], electronic structure [29,32-37, 39,40], photoluminescence and surface photovoltaic effect [41]. In addition, the growth [42, 43] and chemical functionalization [44] of ultrathin InSe nanosheets have been explored, and potential applications based on InSe in nanoelectronics [45], sensors [46], optoelectronics and photodetectors [47-51] have been sought. These studies have revealed many fascinating behaviors of InSe. For example, it possesses a high carrier mobility (up to $10^3$ cm$^2$V$^{-1}$s$^{-1}$ and $10^4$ cm$^2$ V$^{-1}$s$^{-1}$ at room and liquid-helium temperatures, respectively) and a high thermal stability up to 660° C [26,27]. Importantly, InSe sheets are stable under ambient conditions and no decomposition in air is observed [28]. A band gap narrowing of around 0.5 eV was observed from bulk InSe (1.35 eV [29]) to bilayer InSe according to photoluminescence spectroscopy measurement [30,31]. Density functional theory (DFT) calculations revealed that bulk InSe has a direct band gap at Z point [32] and few-layer InSe shows an indirect band gap [33-37]. Despite the presence of heavy atoms of In and Se, the effect of spin-orbital coupling on the states at the top of the valence band and the bottom of the conduction band is negligible, and thus the band gap is less affected by this coupling [32]. Similar to black phosphorus, much lighter carriers are found for transport along the out-of-plane direction than the in-plane direction [32, 34].

For many applications, examining the effects of external factors, such as environment molecules and dopants [52-54], contacting electrodes [55], and supporting substrates on 2D materials [56] is highly important to achieve improved stability, robust performance, and tunable functionality. For instance, water and oxygen molecules were found to affect the stability of phosphorene [57, 58]. Organic molecules were used for liquid exfoliating $MoS_2$ [59] and phosphorene [60]. Modulation of carrier density and polarity of phosphorene [14] and $MoS_2$ [61] were shown by a proper control of defects and molecular functionalization. A recent demonstration of a dye-sensitized photo-sensing device showed that a Lewis acid–base reaction can be used to form planar p-type $[Ti^{4+}_n(InSe)]$ complexes by anchoring InSe layers with $Ti^{4+}$ species [62]. These studies clearly demonstrate the importance of the effects of external factors on 2D materials. We note that there is still lack of knowledge and understanding on the effects of small molecules on the electronic properties of InSe. For practical applications, it is both important and necessary to fill in this knowledge gap about InSe.

In this work, we performed a detailed study on the interaction of InSe with several small molecules, including CO, NO, $NO_2$, $H_2S$, $N_2$, $NH_3$, $O_2$, $H_2O$ and $H_2$, which are ubiquitous in environment. We are particularly interested in the ability of InSe in sensing these molecules. In addition, we would like to reveal the mechanism of charge transfer, which may give valuable hints for developing new strategies for achieving a better structural stability and improved excitonic and optical absorption efficiencies of InSe. Last but not least, we would like to understand the changes in content and polarity of charged carriers and also screening and trapping effects, which may lead to novel optical properties and device applications. Our work indeed reveals many fascinating behaviors of these molecule-adsorbed systems, which show several promising applications, in particular, in gas sensing.

II. **COMPUTATIONAL METHOD**

Our calculations are performed by using Vienna ab initio simulation package (VASP) package [63] within the framework of density functional theory (DFT). For a proper description of the dispersive forces between the molecules and the 2D host, van der Waals (vdW) corrected functional with Becke88 optimization (optB88) is used together with an energy cutoff of 400 eV. The optimized honeycomb lattice constant of the unit cell of monolayer InSe is 4.077 Å. All the atomic positions are fully relaxed until the force on each atom is smaller than 0.005 eV/Å. For modulating the effect of molecular adsorbates in the dilute doping limit, we construct a 4×4 supercell (containing 32 InSe units) and place the molecule above the sheet. To avoid the spurious images interaction, a vacuum layer with thickness of 15 Å is adopted. The Brillouin zone sampling for k-points is based on a 3×3×1 Monkhorst-Pack grid. The adsorption energy ($E_{ad}$) for describing the strength of the interaction between the molecule to the InSe sheet is calculated via $E_{Mol+InSe}$-$E_{InSe}$-$E_{Mol}$, where $E_{Mol}$, $E_{InSe}$ and $E_{Mol+InSe}$ are the energies of the molecule, InSe sheet and molecular adsorbed InSe, respectively.

III. **RESULTS**

**A. Features in the electronic structure of pristine monolayer (1L) InSe**

The band structure of the 4×4 supercell structure of monolayer InSe is shown in Fig. 1a. Our calculation with GGA method shows that monolayer InSe has an indirect band gap of 1.50 eV with the minimum of conduction band being at the Γ point and the valence top sitting in between Γ and M points. These results are consistent with the experimental results that showed limited photoluminescence efficiency of 1L InSe [30]. Note that our predicted band gap is underestimated due to the well-known GGA deficiency. It was shown that pristine InSe samples exhibited a *p*-type conduction [28], consistent with our calculations that the Fermi level ($E_f$) is close to the valence top.

The density of states (DOS) presented in Fig. 1b shows that the bottom of the conduction band mainly comprises the Se-*p* and In-*s* states with slight hybridizations of In-*p* and In-*d* orbitals. The projected charge density at the bottom of the conduction band (right panel in Fig. 1c) shows an anti-bonding character of the electronic states. In contrast, the top of the valence band mainly consists of the In-*p* and lone-pair Se-*p* orbitals with slight In-*d* and In-*s* orbitals. The partial charge analysis of the top valence band (left panel in Fig. 1c) reflects a non-bonding character between the Se-*p* and In-*p* states. In contrast to the lone-pair states of phosphorus distributed far below the $E_f$ in phosphorene, the low-lying lone-pair states of Se is located at the top of the valence band of InSe and close to the $E_f$, implying a high sensitivity to external adsorbates. The weak bonding nature of the top valence states also suggests that the structural stability of this material is dominated by the deep-lying states and structural degradation arising from the depopulation of the valence band is less unlikely.

**B. Molecular donors**

We next examine the adsorption of several typical small molecules (CO, NO, $NO_2$, $H_2S$, $N_2$, $NH_3$ and $H_2$) above the InSe surface. For the adsorption of each molecule, several different configurations with respect to the adsorbing sites are considered including: top of the Se site, top of the In site, and the top site above the center of the hexagonal void. For each site, several relative orientations of the molecules with the sheet are calculated including the vertical and flat configurations for linear molecules and the upward, downward, and tilted alignment of the molecule with the surface. Interestingly, for all the molecules, the most stable adsorption site is at the top site above the center of the hexagonal void.

The oxidation states of the adsorbed molecules are very important in affecting the electronic doping, structural stability, and photoluminescence of InSe. To quantitatively determine the amount of charge transfer, we calculate the differential charge density (DCD) $\Delta\rho(r)$, which is defined as $\Delta\rho(r) = \Delta\rho_{InSe+Mol}(r) - \Delta\rho_{InSe}(r) - \Delta\rho_{Mol}(r)$, where $\Delta\rho_{InSe+Mol}(r)$, $\Delta\rho_{InSe}(r)$, and $\Delta\rho_{Mol}(r)$ are the charge densities of the adsorbed system, isolated InSe sheet and isolated molecule in adsorbing configuration, respectively. The plane-averaged DCD $\Delta\rho(z)$, which is obtained by integrating the $\Delta\rho(r)$ within the InSe x-y basal plane, shows the loss and increase of electrons along the out-of-plane direction. By further integrating the $\Delta\rho(z)$ from bottom infinity to the z point, we can obtain the amount of transferred electrons from the molecule to InSe sheet at z point via $\Delta Q(z) = \int_{-\infty}^{z} \Delta\rho(z') \, dz'$. This allows us to plot the line profiles of $\Delta\rho(z)$ and $\Delta Q(z)$ together, and the value of the $\Delta Q(z_{int})$ at the interface point $z_{int}$ between the adsorbed molecule and InSe gives rise to the exact amount of transferred electrons from the molecule to InSe. The $z_{int}$ is determined by the z point in the gap region where a zero value of $\Delta\rho(z)$ and a maximum of the $\Delta Q(z)$ are located. In this section, we mainly discuss the donor molecules: CO, $H_2$, $H_2S$, and $N_2$.

**CO adsorption:** For the CO adsorption on InSe, the most stable configuration adopts a nearly parallel alignment of the CO molecule to the InSe surface with the CO located at the center of the hexagon around 3.08 Å above the top plane consisting of Se atoms. The $E_{ad}$ is found to be -0.13 eV. The adsorbing configuration and the isosurface plots of the DCD are shown in Fig. 2a. It can be seen that a clear charge accumulation occurs between the C atom and two Se atoms in the vdW's gap formed between the CO and InSe. The $\Delta\rho(z)$ and $\Delta Q(z)$ curves are plotted in Fig. 2b. It can be seen that only a tiny amount of electron transfer, around 0.001 per CO molecule, occurs from CO to InSe. Such a limited charge transfer is consistent with the almost unchanged length of C-O bond of the adsorbed CO molecule compared with the gas molecule.

The local density of states (LDOS) and band structure are shown in Fig. 2c and d, respectively. It can be seen that there is no additional CO induced states within the band gap. Concerning the frontier orbitals like the highest occupied molecular orbital (HOMO) and the lowest unoccupied molecular orbital (LUMO) of the adsorbed CO molecule, its 4σ (HOMO-2), 1π (HOMO-1), 5σ (HOMO) and 2π*(LUMO) states are located at -8.50, -6.23, -3.58 and 3.40 eV relative to the $E_f$. According to the LDOS plot, the 5σ peak, which coincides well with the valence states of the InSe host, is slightly more broadened than 4σ, 1π and 2π* levels. As the 5σ state mainly distributes on the C atom and there are enhanced bonding charges between C and Se atoms as shown in Fig. 2a, it is highly likely that the CO-InSe interaction mainly occurs via the hybridization between the 5σ state and the lone pair electrons on the Se atoms.

**H$_2$ adsorption:** Similar to the case of CO molecule, the lowest energy configuration of H$_2$ above InSe also adopts a parallel alignment with the basal plane (Fig. 3a). The H$_2$ is located at about 3.01 Å above the terminated Se lattice plane with the length of H-Se bonds ranges from 3.60 to 3.77 Å. The corresponding $E_{ad}$ is -0.05 eV, comparable to that of H$_2$ above graphene with $E_{ad}$ of -0.04 eV/H$_2$ [64]. It is well known that graphene is a potential material for hydrogen storage due to its appropriate H$_2$ binding energy for simultaneously stable hybrogen storage and facile release [65]. Herein the predicted comparable adsorption energy of H$_2$ above InSe suggests that it may be a promising material for hydrogen storage despite its smaller gravimetric value due to its larger density than graphite.

Charge transfer analysis shows that there is a strong depletion of electrons in the H$_2$ molecule upon adsorption. The isosurface plot of the DCD shown in Fig. 3a reveals that diminishing electrons mainly distribute at the two ends of H$_2$. An increase of electrons at the nearest Se atoms can be clearly observed. Charge integration analysis is presented in Fig. 3b and around 0.146 electrons are transferred from each H$_2$ molecule to the InSe sheet. Interestingly, while the H$_2$ molecule donates more electrons to InSe than the adsorbed CO, its $E_{ad}$ is much smaller than that of CO. This may be attributed to the dominated dipole-dipole interaction which is responsible for the physisorbed CO above InSe. The dispersive forces between the heteropolar CO molecule and InSe should be much stronger than that of the zero-dipole moment H$_2$ molecule. The LDOS and band structure plots (Fig. 3c and d) show that there is no hydrogen induced states within the band gap. However, the $E_f$ is now shifting from the band gap to the conduction band in the $H_2$ adsorbed InSe, indicating a n-type conduction with a strong electron donating ability of $H_2$ molecule. Such a strong charge transfer causes a spin splitting of the $H_2$ level which is located around 8 eV below $E_f$ (Fig. 3c).

**$H_2S$ adsorption:** The lowest-energy configuration of the adsorption of $H_2S$ adopts a geometry with the H-S-H plane aligned parallel to the InSe sheet (Fig.4a). The two S-H bonds point toward the In atoms, respectively and the molecule is located at above 2.92 Å above the Se basal plane. $H_2S$ molecule is a Lewis type molecule which has six valence electrons with two forming high lying σ bonds and two low-energy orbitals occupying two sets of lone-pairs electrons. These lone-pair electrons couple strongly with the lone-pair electrons of surface Se atoms in InSe with the $E_{ad}$ of -0.21 eV. The DCD analysis shows that the $H_2S$ loses electrons while the InSe receives electrons. Significant amount of electrons is donated to the nearest Se atoms close to the molecule. A tiny amount of electron transfer can be found in the second nearest Se atoms in the top surface, which should be induced by the relatively large dipole moment of $H_2S$. The $\Delta\rho(z)$ and $\Delta Q(z)$ curves plotted in Fig. 4b show that around 0.016 electrons are donated by each $H_2S$ molecule.

The DOS and band structures are shown in Fig. 4c and d, respectively. No $H_2S$ induced states are found within the band gap. The three highest occupied levels ($1b_2$, $3a_1$ and $1b_1$) of $H_2S$ are marked in the DOS plot. Different from the CO and H2 cases, these orbitals are completely located within the valence bands of InSe. The peaks of these levels are greatly broadened, indicating a strong hybridization of InSe despite a physisorption above the InSe surface. This suggests that the $H_2S$ molecule may significantly affect the electronic performance of InSe. The $1b_1$ level of the electrons, which are distributed within the two S-H bonds and facilitate the orbital mixing with the underlying Se atoms, shows the largest broadening. The band gap reduces from 1.50 eV of pure InSe to 1.43 eV upon adsorbing $H_2S$ (Table 1).

**$N_2$ adsorption:** The most stable configuration of the adsorbed $N_2$ molecule adopts a nearly parallel configuration relative to the basal plane. It is located at around 2.90 Å above the center of the hexagonal void. The $E_{ad}$ is calculated to be -0.12 eV. Figure 5a shows the isosurface of DCD. It can be seen that only a tiny amount of charge transfer occurs between the $N_2$ molecule and InSe, which mainly involves the nearest neighboring Se atoms. This tiny charge transfer can be attributed to the inert nature and the zero-dipole moment of the homopolar $N_2$ molecule. From the $\Delta\rho(z)$ and $\Delta Q(z)$ curves as shown in Fig. 5b), there is an amount of around 0.005 electrons per $N_2$ molecule transferred to the InSe sheet. Owing to the strong triple nitrogen bond and the inert nature of the $N_2$ molecule, it is reasonable that there are no additional hybridized states in the DOS and band structure for the $N_2$ adsorbed InSe system (Fig. 5c and d).

**C. Molecular acceptors**

**$NH_3$ adsorption:** The lowest-energy configuration for the $NH_3$ adsorbed InSe system is shown in Fig. 6a, where the molecule is located at above the hollow hexagon center with the N atom pointing toward the surface and the three H atoms pointing away from the surface (Fig. 6a). The $NH_3$ molecule is about 2.49 Å above the Se basal plane and the length of the three N-Se bonds ranges from 3.60 to 3.70 Å. The $E_{ad}$ is found to be -0.20 eV. Figure 6b shows that there is a significant charge redistribution between the $NH_3$ and InSe. While there is a clear charge accumulation in the Se atoms, a strong charge depletion occurs in the vdW gap between the molecule and InSe. The $\Delta\rho(z)$ and $\Delta Q(z)$ curves shown in Fig. 6b indicate that around 0.019 electrons are transferred from InSe to $NH_3$. Interestingly, the oxidation state of $NH_3$ molecule above InSe is different from the adsorption above phosphorene surface, where $NH_3$ molecule is a charge donor [14] although the adsorption configurations for the two cases are similar and both 2D materials have lone-pair electrons in surface atoms.

It should be noted that the way of charge transfer in the case of $NH_3$ is slightly different from the cases of preceding donor molecules. While clear lobes and nodal planes appear in the vdW gap of the DCD plots in the donor molecules like CO, $H_2$, $H_2S$ and $N_2$, no such characteristics appears in the DCD plot of NH3 in Fig. 6a. It seems that the charges are redistributed via forming slight hybridized states between the $NH_3$ and InSe. The DOS plot (Fig. 6c) shows that the nonbonding HOMO orbital ($3a_1$), which consists of the lone electron pair at the N atom, is located at around -0.97 eV below the $E_f$. The large broadening of this level could be an indicator of its hybridization with valence states of InSe. Instead, the doubly degenerated $NH_3$1e level, which consists of the s-p hybridized molecular orbitals, is located separately with respect to the valence bands of InSe, indicating a relatively weak interaction. Similar to other molecular dopants, there are negligible changes in the curves of the band structure (Fig. 6d) due to $NH_3$ adsorption. However, the band gap reduces from 1.50 eV of pure InSe to 1.45 eV upon adsorbing $NH_3$ (Table 1).

**NO adsorption:** For the NO molecule, its adsorbing geometry and energetics are similar to those of CO adsorption. The most stable adsorbing configuration shows the NO lying parallel to the surface with 3.10 Å above the center of the hexagon and the $E_{ad}$ of -0.13 eV, comparable to that of CO (see Table 1). However, the NO molecule is an open-shell molecule and shows a very different charge transfer and electronic modification of InSe than that of CO adsorption. According to the DCD plot shown in Fig. 7a, there exist typical characters of redistribution of the electrons in the NO molecule reflected by the orbital-like lobes of the diminishing and accumulating electronic densities. This suggests that the population of some NO orbitals is increased with some other orbitals and becomes less occupied upon contacting with InSe. At the InSe side, most of transferred charges are distributed at the closest Se atoms, suggesting that the interacting mechanism involves the bonding of the lone-pair electrons of Se atoms with the valence electrons of NO. The exact charge transfer analysis (Fig. 7b) reveals that around 0.094 electrons are transferred from InSe to NO. The acceptor role of NO on InSe is similar to that of the NO on phosphorene [14], but different from that of NO on graphene, where NO serves as a donor[66].

Free NO gas molecule has a singly occupied electron in a doubly degenerated antibonding $1\pi^*$ orbital, followed by a lower-energy doubly degenerated $1\pi$ orbital and a $2\sigma$ orbital. As can been from the DOS and band structure shown in Fig. 7c and d, the state hybridization and charge transfer between NO and InSe strongly alter the degeneracy of orbitals of NO. All the levels become spin-split, leading to a magnetic moment of 1 $\mu_B$ of the adsorbed system. In addition, the degeneracy of $1\pi^*$ orbital is lifted and split into two levels located within the band gap at around 0.10 and 0.25 eV below the conduction band of InSe. The presence of the molecularly induced states could modify the optical properties and photoluminescence of InSe as NO could be a potential trapping center.

**$NO_2$ adsorption:** The most stable configuration of the adsorbed $NO_2$ on InSe has a tilted geometry with the $NO_2$ being located at about 2.71 Å above the hexagon center of the top Se basal plane, with the two O atoms pointing toward the In atoms (Fig. 8a). The $NO_2$ has the strongest adsorption with the $E_{ad}$ of -0.24 eV among all the considered molecules, which could arise from the coexistence of a large dipole moment of $NO_2$ and resonant molecular levels with the InSe states. As shown in the isosurface plot of DCD (Fig. 8a), a clear accumulation of electrons in the adsorbed $NO_2$ molecule and a loss of electrons in the Se atoms can be observed. This indicates that $NO_2$ has a relatively big ability to withdraw electrons from InSe. The total amount of electrons received per $NO_2$ molecule is found to be 0.039 electrons judging from the $\Delta\rho(z)$ and $\Delta Q(z)$ curves as shown in Fig. 8b.

The state alignment and hybridization between $NO_2$ and InSe can be reflected in the plots of the DOS and band structure (Fig. 8c and d). The $6a_1$ orbital of $NO_2$ becomes spin split and the HOMO state ($6a_1$, spin-up) is around 0.90 eV below $E_f$ while the LUMO ($6a_1$, spin-down) state is around 0.40 eV above the $E_f$, giving rise to a total magnetic moment of $1\mu_B$. The $4b_1$ and $1a_2$ $NO_2$ orbitals coincide with the valence states of InSe. The orbital mixing and hybridization should account for the charge transfer between $NO_2$ and InSe shown above. In addition, similar to the NO case, the presence of $NO_2$ molecule induced in-gap state ($6a_1$, spin down) could potentially alter the optical properties of InSe. The band gap reduces from 1.50 eV of pure InSe to 1.45 eV upon adsorbing $NO_2$ (Table 1).

**$O_2$ and $H_2O$ adsorption:** The $O_2$ and $H_2O$ molecules are the two common molecules in air which can greatly affect the structural stability of the 2D materials [67-71]. It is well-known that the two molecules are the direct cause of the structural degradation of phosphorene when exposed in air. For the physisorption of the $O_2$ and $H_2O$ above InSe, the charge transfer and electronic properties of the adsorbed systems are shown in Fig. 9. Concerning the adsorbed $O_2$ molecule, the most stable configuration shows a flat alignment of the $O_2$ molecule relative to the InSe basal plane with locating around 3.19 Å above the hexagon center of the top Se atomic plane. In contrast to its significant adsorption and oxidizing ability above phosphorene sheet, the $O_2$ molecule has relatively weak $E_{ad}$ (-0.12 eV) and negligible charge transfer (-0.001 e per $O_2$) with InSe sheet. Therefore, the InSe sheet may hard to be oxidized at ambient condition. The plot of band structure (Fig. 9c) shows that the antibonding LUMO state ($2\pi$, down) is located in the band gap of InSe. For the $H_2O$ molecule, the most stable adsorbing configuration shows that $H_2O$ molecule locates around 2.37 Å above at the hexagon center of the top Se atomic plane with the H-O-H plane aligning vertical to the InSe plane and one of the H-O bonds pointing away the normal direction of the surface. The water molecule is found to accept electrons (around 0.01 e per molecule) from the InSe surface. Therefore, the water molecule is an oxidizing acceptor molecule above the InSe sheet, different from its role (molecular donor) in phosphorene.

## IV. DISCUSSION

The recent experiments demonstrated a high mobility of InSe [26,27]. Understanding the effects of environmental molecules on its electronic and chemical properties should be highly important for its potential applications. Figure 9 shows the relationship of the dipole moment and $E_{ad}$ for the adsorbed molecules on the InSe surface. For comparison, we also plot the data for the molecular adsorption on the phosphorene with the same computational method [14]. For both cases, the curve adopts a "**V**" shape. Starting from the zero dipole molecules, the adsorption enhances with increasing the dipole moment of the adsorbates, reaches maximum for the $NO_2$ molecule and then weakens afterward with further increasing the dipole moment. Two mechanisms are responsible for the interaction of the molecules with the surface: chemical hybridization and the dipole-dipole interaction. It is expected that the former mechanism is dominated for adsorptions of molecules with a small dipole moment. As shown in Fig. 9, the $E_{ad}$ for the adsorption of the small-moment molecules ($H_2$, CO,

NO, $NO_2$) on InSe is around a half of those for adsorption on phosphorene. This suggests that the chemical interaction between the molecule and InSe is weaker than that between the molecule and phosphorene, which is well-known for its sensitivity to environmental molecules and low structural degradation. The overall much weaker $E_{ad}$ for molecules on the InSe surface suggests that InSe in air should be more stable than phosphorene in air.

Our calculations have shown that the carrier density and polarity of the InSe sheet can be easily modulated upon selective adsorption of gas molecules. It should be noted that in our current study, we only consider the dilute doping of molecular adsorbates. In high doping condition, the strength of interaction and the degree of charge transfer per molecule should be smaller than our predicted values due to the increasingly important role of the intermolecular interaction. While our current work only investigates the gas adsorption on monolayer InSe, it is expected that the predicted oxidation states of these molecules should remain true for few-layer InSe since the surface chemistry is similar and the interlayer interaction is through weak vdW interaction.

In addition, the charge transfer induced by the molecular adsorption may allow the modulation of optical properties of InSe. Due to the atomically thin structure of InSe, the electronic screening effect is expected to be weak and its optical properties may exhibit a high sensitivity to environment. Previous studies show that the amount of charge transfer induced by chemical molecules can greatly suppress the non-radiative recombination in 2D transition metal dichalcogenides [72-75] and in some cases, it may even lead to near-unity photoluminescence [76]. The photoluminescence can also be modulated by controlling the charge transfer by selecting different substrates [77]. For InSe, the indirect band gap of monolayer and few-layer InSe should have a weak efficiency of photoluminescence. However, the presence of direct band gap in bulk InSe may allow a strong emission of excitons. Investigating the thickness-dependent evolution of photoluminescence under the exposure to different gas environment, surface treatment or substrates should be highly interesting and important.

Another important quantity that may be significantly affected by charge transfer in InSe is the band bending near the surface or the interface region. As a p-type conduction, the $E_f$ of InSe is close to the valence band and a downward band bending is highly likely when contacting with metal electrodes or other semiconducting materials due to the spilling of the hole carriers. It is expected that the gas adsorption of molecules can tailor the band bending due to a different accumulation of carriers and modifying the built-in potential. Through modifying the electric field, the additional charge transfer induced by the adsorbed molecules can either enhance or suppress the electron-hole recombination in InSe, depending on the oxidation states of the adsorbed molecule.

Since InSe shows a very good light adsorbing ability at visible light region and a strong photovoltaic, the response of physical quantities, such as band bending and photoluminescence with light excitation and gas treatment, would be highly interesting. This measurement allows a quantitative estimation of charge transfer induced by the molecules. For instance, the photon-stimulated desorption method, which was commonly used for monitoring the desorption behavior of $O_2$ molecule on $TiO_2$ surface [78-81], can be effectively used for probing the charge transfer and the variation of hole states near the surface due to the incident photons. Other methods such as IR/Raman measurements can also help to determine the charged states of the adsorbed molecules through comparing the frequencies of adsorbed chemical species with those of their free states [82].

## V. CONCLUSION

We investigated the energetics and charge transfer of several small molecules physisorbed on monolayer InSe using first-principles calculations. Our calculations revealed that the lone-pair states of Se were located at the top of the valence band of InSe and close to the Fermi energy level, implying strong interacting ability of InSe with external molecules. We found that the $NH_3$, $NO_2$, $H_2O$ and NO molecules were strong acceptors whereas $H_2S$ and $H_2$ were strong donors. The strong charge donating ability of $H_2$ could even lead to a semiconductor-metal transition of InSe with a single $H_2$ adsorbing in the supercell. Interestingly, the $NH_3$ molecule was found to be a strong acceptor, which is different from its well-known character as a molecular donor on phosphorene and graphene. Such significant modification of electronic properties together with an appropriate adsorption strength implies promising applications of InSe as gas sensors.

**AUTHOR INFORMATION**

**Corresponding Author**

zhangg@ihpc.a-star.edu.sg; zhangyw@ihpc.a-star.edu.sg

**Notes**

The authors declare no competing financial interests.

**ACKNOWLEDGMENT**

This work was supported in part by a grant from the Science and Engineering Research Council (152-70-00017). The authors gratefully acknowledge the financial support from the Agency for Science, Technology and Research (A*STAR), Singapore and the use of computing resources at the A*STAR Computational Resource Centre, Singapore.

Table 1. The adsorption energy ($E_{ad}$), the amount of charge transfer ($\Delta q$), and the height ($h$) from the molecule to the Se plane, the donor/acceptor characteristics of the molecular dopant on the InSe surface, and the band gap ($E_g$) of adsorbed systems. Note that a positive (negative) $\Delta q$ indicates a loss (gain) of electrons from each molecule to InSe.

| Molecules | $E_{ad}$(eV) | $\Delta q$ (e) | $h$ (Å) | Molecule on InSe | $E_g$ (eV) |
|---|---|---|---|---|---|
| CO | -0.13 | 0.001 | 3.08 | -- | 1.49 |
| H$_2$ | -0.05 | 0.146 | 3.01 | donor | 1.49 |
| H$_2$S | -0.21 | 0.016 | 2.92 | donor | 1.43 |
| N$_2$ | -0.12 | 0.005 | 2.90 | -- | 1.49 |
| NH$_3$ | -0.20 | -0.019 | 2.49 | acceptor | 1.45 |
| NO | -0.13 | -0.094 | 3.10 | acceptor | 1.49 |
| NO$_2$ | -0.24 | -0.039 | 2.71 | acceptor | 1.45 |
| O$_2$ | -0.12 | -0.001 | 3.19 | -- | 1.50 |
| H$_2$O | -0.17 | -0.01 | 2.37 | acceptor | 1.49 |

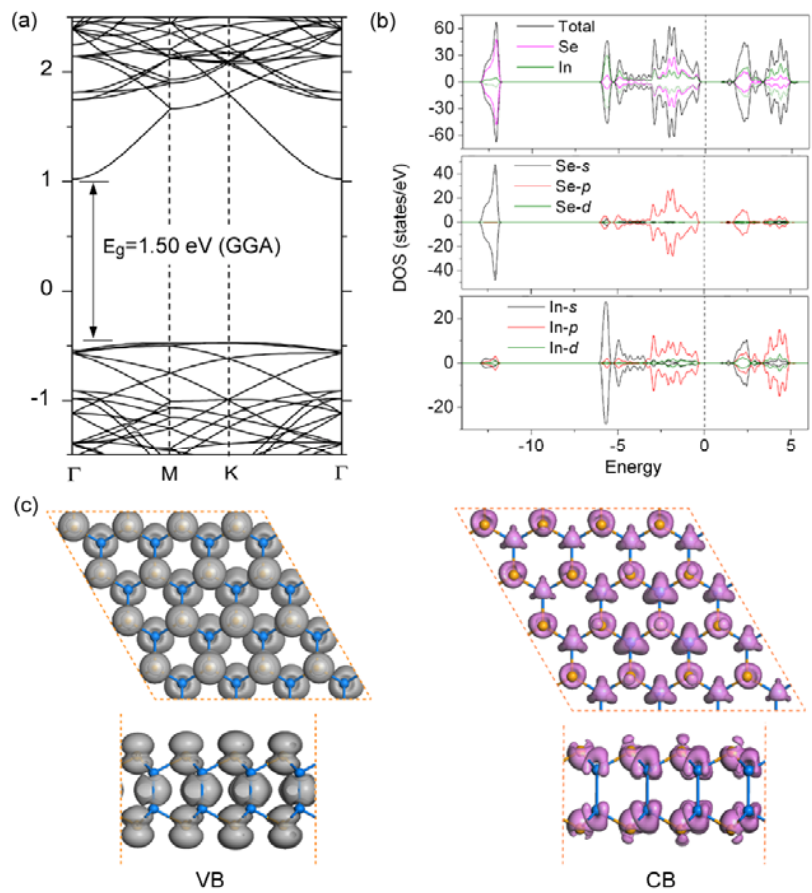

Fig. 1. (a) Band structure of monolayer InSe with GGA calculations. (b) DOS and charge distribution of the valence band (VB) top and the conduction band (CB) bottom of monolayer InSe.

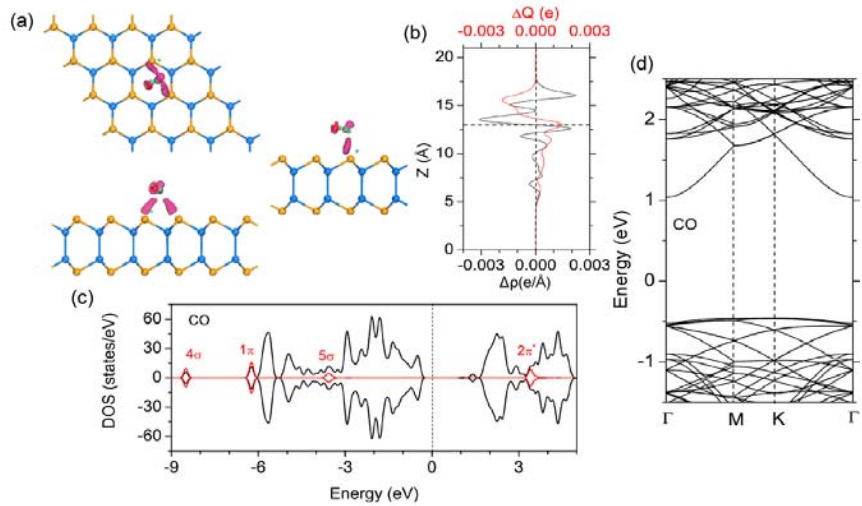

Fig. 2. CO adsorbed on monolayer InSe. (a) The top and side views of lowest-energy configuration of CO on InSe. The charge transfer between CO and InSe is illustrated by the isosurface (0.001 Å$^{-3}$) plots of the differential charge density, where a red (green) color represents an accumulation (loss) of electrons. (b) The line profiles of the plane-averaged differential charge density $\Delta\rho(z)$ (black line) and the transferred amount of charge $\Delta Q(z)$ (red line). (c) Total DOS (black line) and LDOS of CO molecule, which are enlarged by a factor of 2. (d) Band structure.

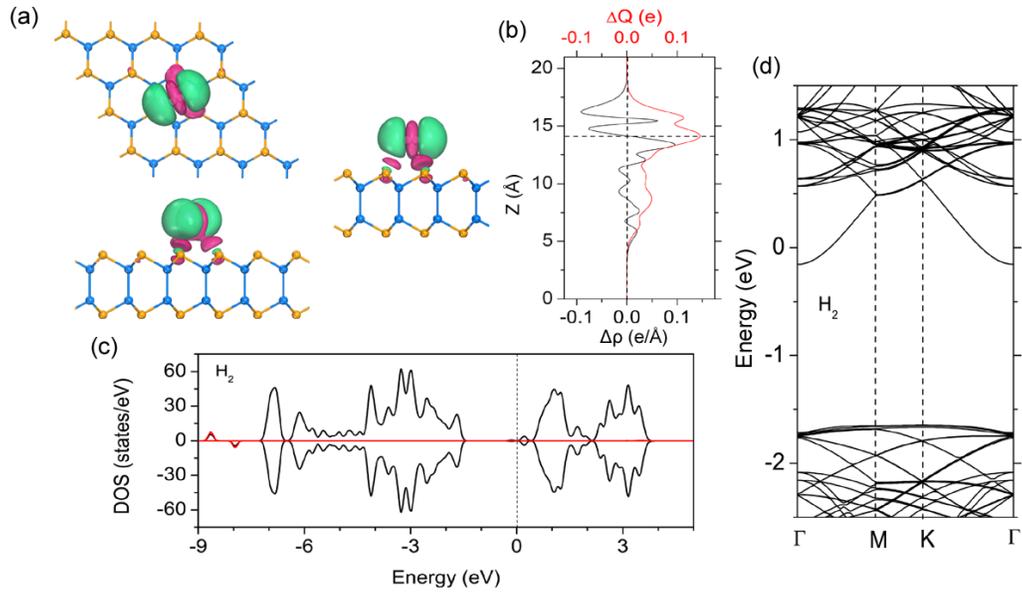

Fig. 3. $H_2$ adsorbed on monolayer InSe. (a) The top and side views of lowest-energy configuration of $H_2$ on InSe. The charge transfer between $H_2$ and InSe is illustrated by the isosurface (0.002 Å$^{-3}$) plots of the differential charge density, where a red (green) color represents an accumulation (loss) of electrons. (b) The line profiles of the plane-averaged differential charge density $\Delta\rho(z)$ (black line) and the transferred amount of charge $\Delta Q(z)$ (red line). (c) Total DOS (black line) and LDOS of $H_2$ molecule, which are enlarged by a factor of 2. (d) Band structure.

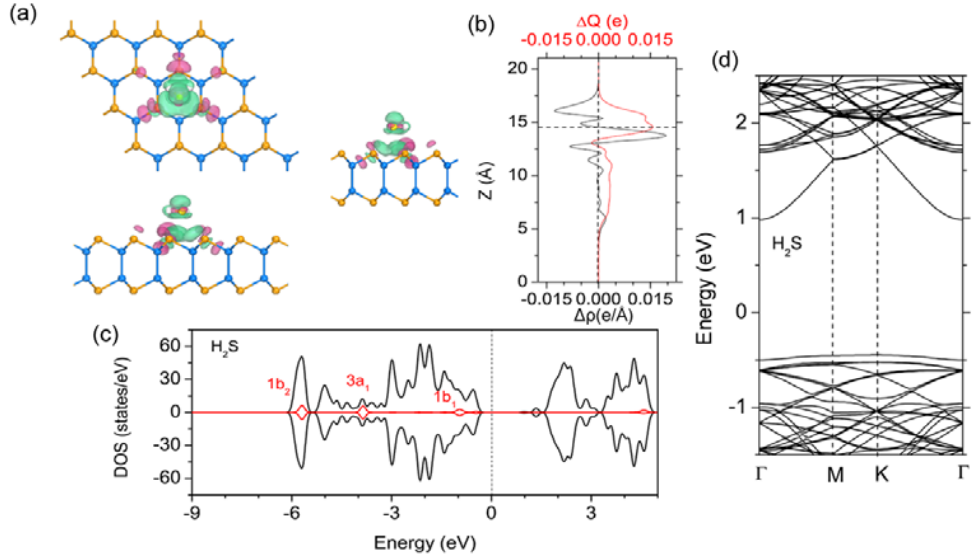

Fig. 4. $H_2S$ adsorbed on monolayer InSe. (a) The top and side views of lowest-energy configuration of $H_2S$ on InSe. The charge transfer between $H_2S$ and InSe is illustrated by the isosurface (0.001 Å$^{-3}$) plots of the differential charge density, where a red (green) color represents an accumulation (loss) of electrons. (b) The line profiles of the plane-averaged differential charge density $\Delta\rho(z)$ (black line) and the transferred amount of charge $\Delta Q(z)$ (red line). (c) Total DOS (black line) and LDOS of $H_2S$ molecule, which are enlarged by a factor of 2. (d) Band structure.

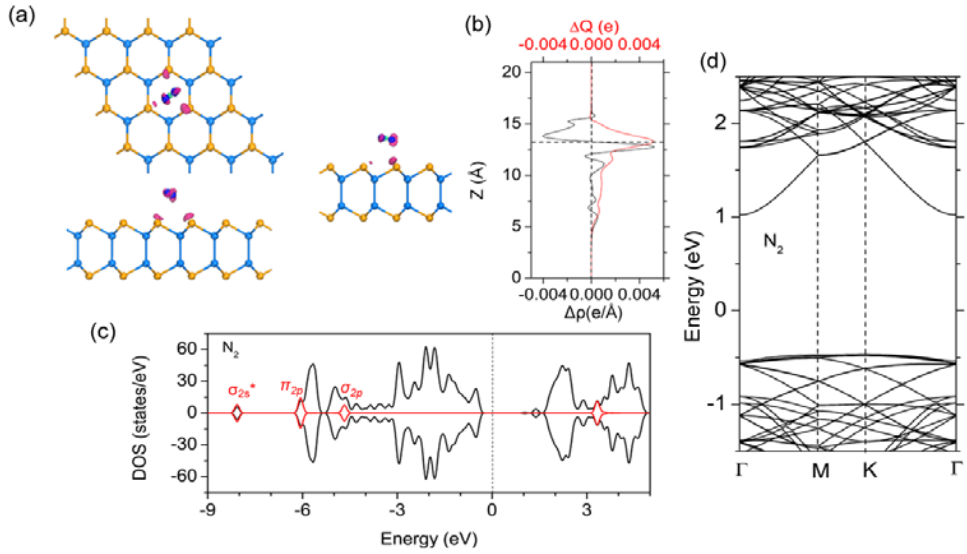

Fig. 5. $N_2$ adsorbed on monolayer InSe. (a) The top and side views of lowest-energy configuration of $N_2$ on InSe. The charge transfer between $N_2$ and InSe is illustrated by the isosurface (0.001 Å$^{-3}$) plots of the differential charge density, where a red (green) color represents an accumulation (loss) of electrons. (b) The line profiles of the plane-averaged differential charge density $\Delta\rho(z)$ (black line) and the transferred amount of charge $\Delta Q(z)$ (red line). (c) Total DOS (black line) and LDOS of $N_2$ molecule, which are enlarged by a factor of 2. (d) Band structure.

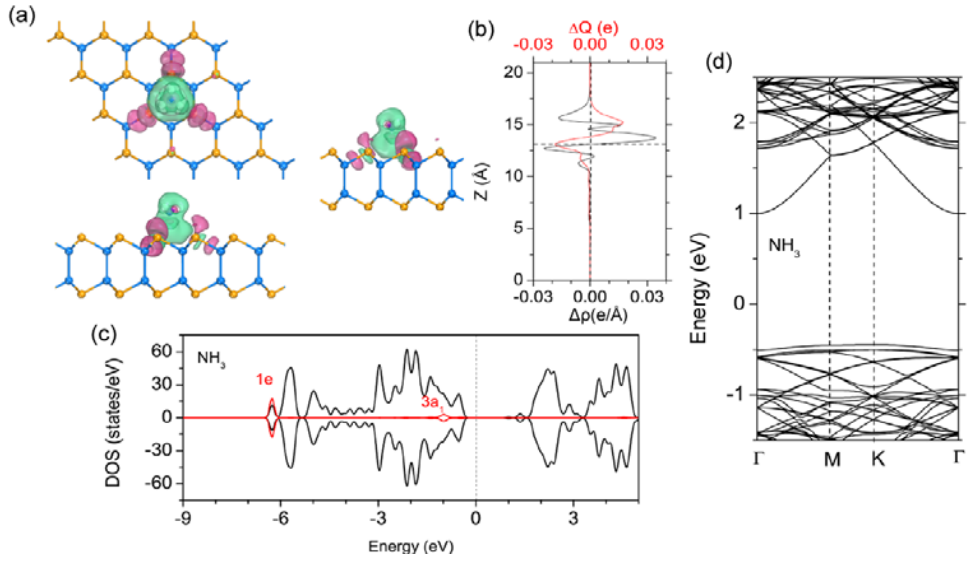

Fig. 6. NH$_3$ adsorbed on monolayer InSe. (a) The top and side views of lowest-energy configuration of NH$_3$ on InSe. The charge transfer between NH$_3$ and InSe is illustrated by the isosurface (0.001 Å$^{-3}$) plots of the differential charge density, where a red (green) color represents an accumulation (loss) of electrons. (b) The line profiles of the plane-averaged differential charge density Δ$\rho$(z) (black line) and the transferred amount of charge Δ$Q$(z) (red line). (c) Total DOS (black line) and LDOS of NH$_3$ molecule, which are enlarged by a factor of 2. (d) Band structure.

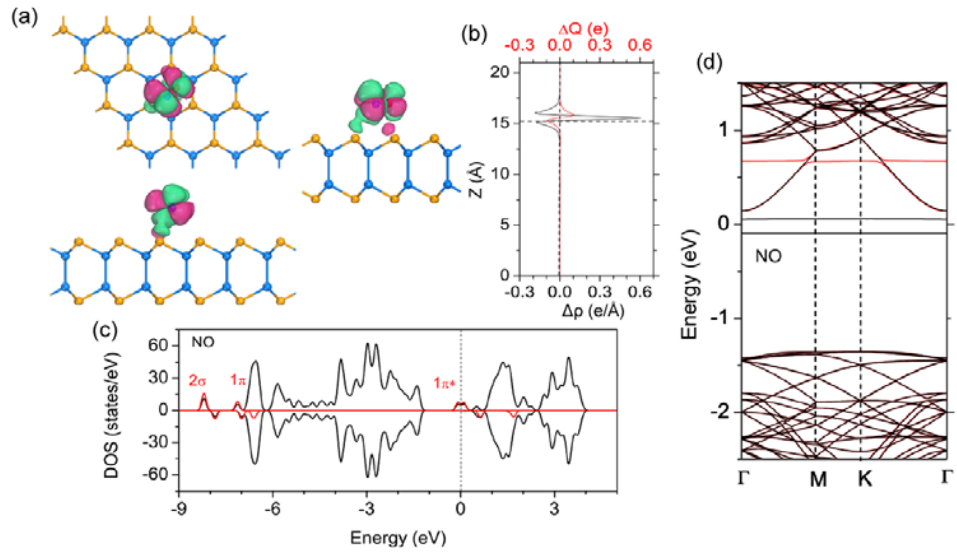

Fig. 7. NO adsorbed on monolayer InSe. (a) The top and side views of lowest-energy configuration of NO on InSe. The charge transfer between NO and InSe is illustrated by the isosurface (0.001 Å$^{-3}$) plots of the differential charge density, where a red (green) color represents an accumulation (loss) of electrons. (b) The line profiles of the plane-averaged differential charge density $\Delta\rho(z)$ (black line) and the transferred amount of charge $\Delta Q(z)$ (red line). (c) Total DOS (black line) and LDOS of NO molecule, which are enlarged by a factor of 2. (d) Band structure.

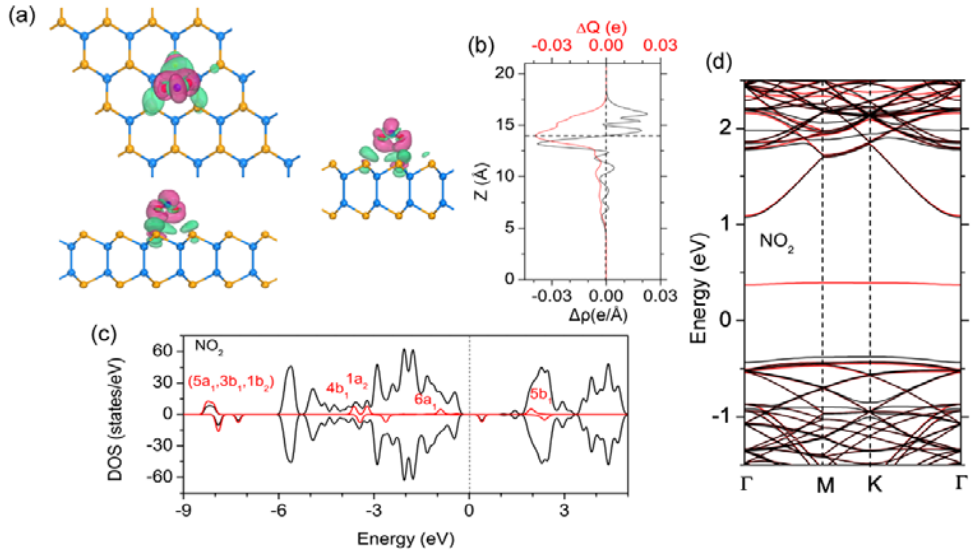

Fig. 8. $NO_2$ adsorbed on monolayer InSe. (a) The top and side views of lowest-energy configuration of $NO_2$ on InSe. The charge transfer between $NO_2$ and InSe is illustrated by the isosurface (0.001 Å$^{-3}$) plots of the differential charge density, where a red (green) color represents an accumulation (loss) of electrons. (b) The line profiles of the plane-averaged differential charge density $\Delta\rho(z)$ (black line) and the transferred amount of charge $\Delta Q(z)$ (red line). (c) Total DOS (black line) and LDOS of $NO_2$ molecule, which are enlarged by a factor of 2. (d) Band structure.

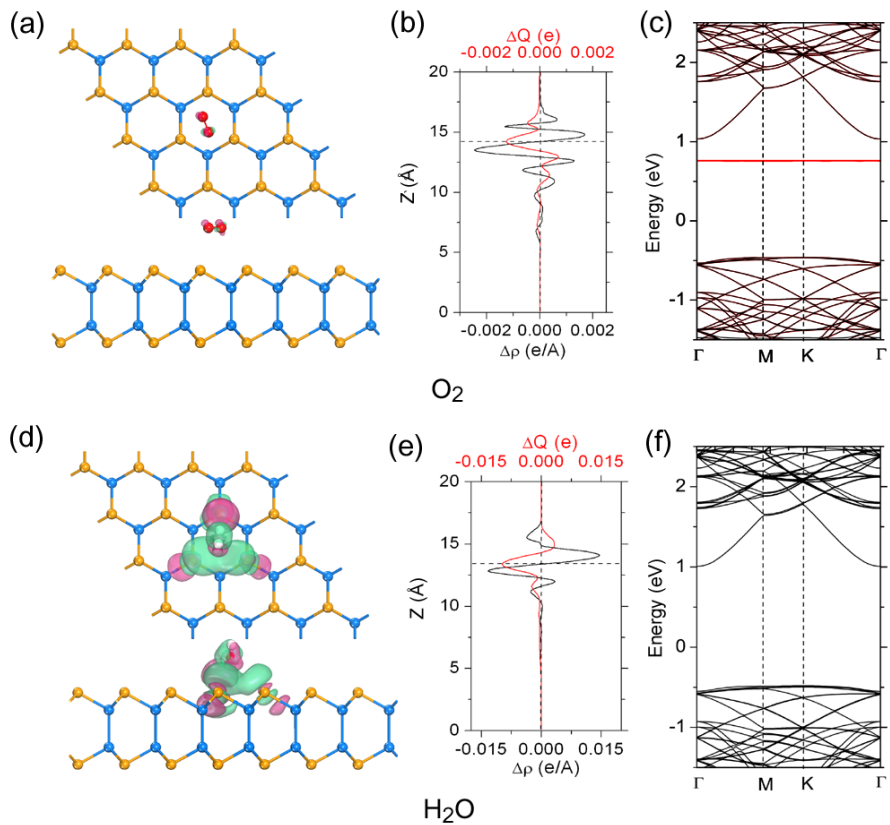

Fig. 9. $O_2$ (a-c) and $H_2O$ (d-f) adsorbed on monolayer InSe. (a,d) The top and side views of lowest-energy configuration of the molecule on InSe. The charge transfer between the molecule and InSe is illustrated by the isosurface (0.001 Å$^{-3}$) plots of the differential charge density, where a red (green) color represents an accumulation (loss) of electrons. (b, e) The line profiles of the plane-averaged differential charge density $\Delta\rho(z)$ (black line) and the transferred amount of charge $\Delta Q(z)$ (red line). (c,f) Band structure of molecular adsorbed InSe. The black and red lines correspond to the spin up and down components respectively.

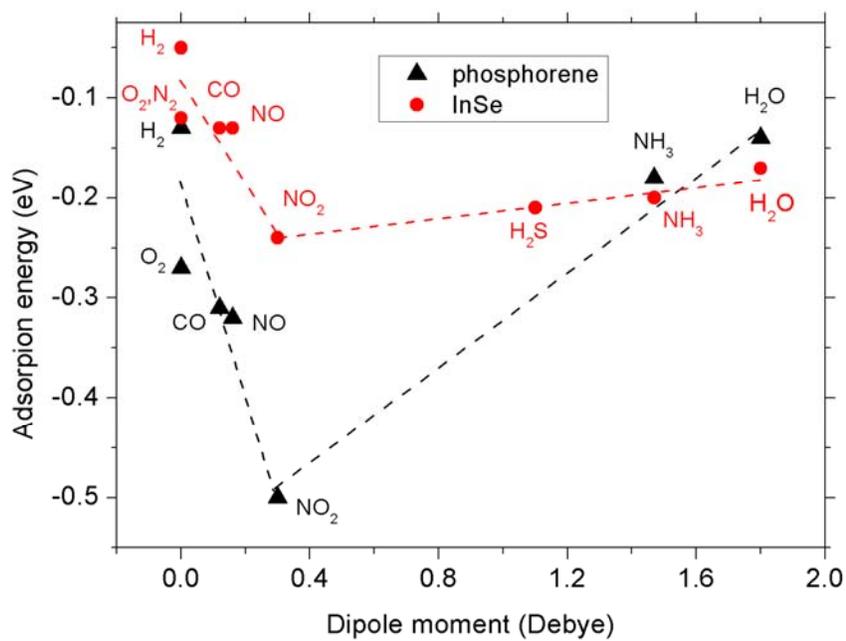

Fig. 10. Comparison of the dipole moment vs. adsorption energy for different molecules physisorbed on InSe and phosphorene. The data of gas molecules adsorbed on phosphorene is adapted from Ref. [14] with the same computational method.

TOC

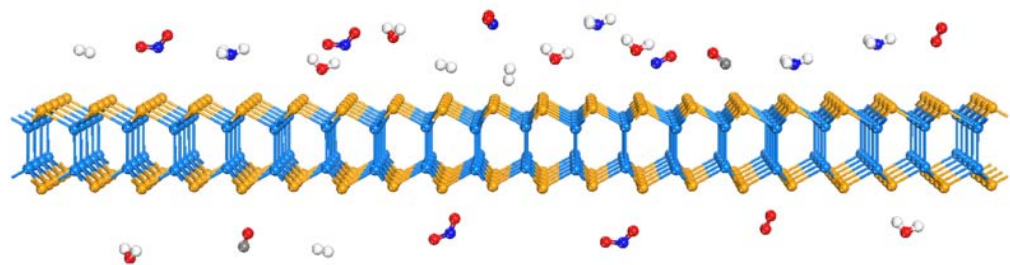